\documentclass[%
 prc,reprint,
showpacs,preprintnumbers,
 amsmath,amssymb,
 aps,
floatfix
]{revtex4-1}
\usepackage{graphicx}
\usepackage{dcolumn}
\usepackage{bm}
\usepackage{epstopdf}
\usepackage{hyperref}

\newcommand{\comment}[1]{}

\newcommand{\non}{\nonumber	\\}

\newcommand{\bra}[1]{\langle #1 |}
\newcommand{\ket}[1]{| #1 \rangle}

\def\mQQ{Q\bar{Q}}
\def\QQ{$\mQQ$}
\def\mbb{b\bar{b}}
\def\bb{$\mbb$}
\def\muns{\Upsilon(nS)}
\def\mus{\Upsilon(1S)}
\def\muss{\Upsilon(2S)}
\def\musss{\Upsilon(3S)}
\def\uns{$\muns$}
\def\us{$\mus$}
\def\uss{$\muss$}
\def\usss{$\musss$}
\def\mcnp{\chi_b(nP)}
\def\mcp{\chi_b(1P)}
\def\mcpp{\chi_b(2P)}

\def\cnp{$\mcnp$}
\def\cp{$\mcp$}
\def\cpp{$\mcpp$}

\def\mRprel{R^\text{prel}_{AA}}
\def\Rprel{$\mRprel$}
\def\del{\partial}

\def\mGammaTot{\Gamma_\text{tot}}
\def\mGammaDamp{\Gamma_\text{damp}}
\def\mGammaDiss{\Gamma_\text{diss}}
\def\GammaTot{$\mGammaTot$}
\def\GammaDamp{$\mGammaDamp$}
\def\GammaDiss{$\mGammaDiss$}

\allowdisplaybreaks

\begin{document}

\preprint{APS/123-QED}

\title{\texorpdfstring{$\Upsilon$}{} suppression
in PbPb
collisions at the LHC}

\author{Felix Nendzig}
\author{Georg Wolschin}%
 \email{wolschin@uni-hd.de}
\affiliation{%
 Institut f{\"ur} Theoretische
Physik
der Universit{\"a}t Heidelberg, Philosophenweg 16, D-69120 Heidelberg, Germany, EU\\
}%

\date{\today}

\begin{abstract}
We suggest that the combined effect of screening, gluon-induced dissociation, collisional damping, and reduced feed-down
explains most of the sequential suppression of \uns\ states that has been observed in PbPb relative to $pp$ collisions
at  $\sqrt{s_{NN}}$ = 2.76 TeV. The suppression is thus a clear, albeit indirect, indication for the presence of a QGP. The \us\ ground state suppression is essentially due to reduced feed-down, collisional damping and gluodissociation, whereas screening prevails for the suppression of the excited states.
\end{abstract}

\maketitle
\section{\label{sec:intro}Introduction\protect}
The suppression of quarkonium (\QQ) states is one of the most promising probes for the properties of the quark-gluon
plasma (QGP) that is generated in heavy-ion collisions at highly relativistic energies. In the QGP the confining
potential is screened due to the interaction of the heavy \QQ\ with medium partons and hence, charmonium and bottomium
states successively melt \cite{ms86} at sufficiently high temperatures $T_\text{diss}$ beyond the critical value $T_c
\approx 170$ MeV.

However, additional processes such as gluon-induced dissociation, and collisional damping contribute to the suppression,
and are effective in a temperature region where the $\Upsilon(nS)$ states -- and in particular, the
$\Upsilon(1S)$ ground state -- have not yet melted due to screening.

Here we concentrate on such processes. It turns out that in particular for the $\Upsilon(1S)$ ground state, bottomium dissociation is not just static screening, but mostly caused by other means -- whereas the dissociation of the excited states is essentially due to screening.

Charmonium suppression has been studied since 1986 in great detail both theoretically
\cite{patra-srivastava-2001,kluberg-satz-2009,kharzeev-2007}, and experimentally at energies reached at the CERN
SPS, BNL RHIC \cite{at09}, and the CERN LHC \cite{alice-2011,silvestre-2011}.
Bottomium suppression is expected to be a cleaner probe. The \us\ ground state with mass $9.46$ GeV is strongly bound. It melts as
the last \QQ\ in the QGP (depending on the potential) only at about $4.10$ $T_c$ \cite{wong2005}. Even at LHC energies the number of
$b\bar{b}$-pairs in the QGP remains small such that statistical recombination is unimportant.

$\Upsilon$ suppression in heavy-ion collisions
has recently been observed for the first time both by the STAR experiment at RHIC \cite{STAR-2011}, and by the CMS experiment at LHC
\cite{CMS2011a,CMS2012b}. CMS data from the 2011 run \cite{CMS-2012} have much better statistics such that the \uss\ state can now
be resolved individually in PbPb collisions at the LHC.

\begin{figure}[tph]
\centering
\includegraphics[width=8.6cm]{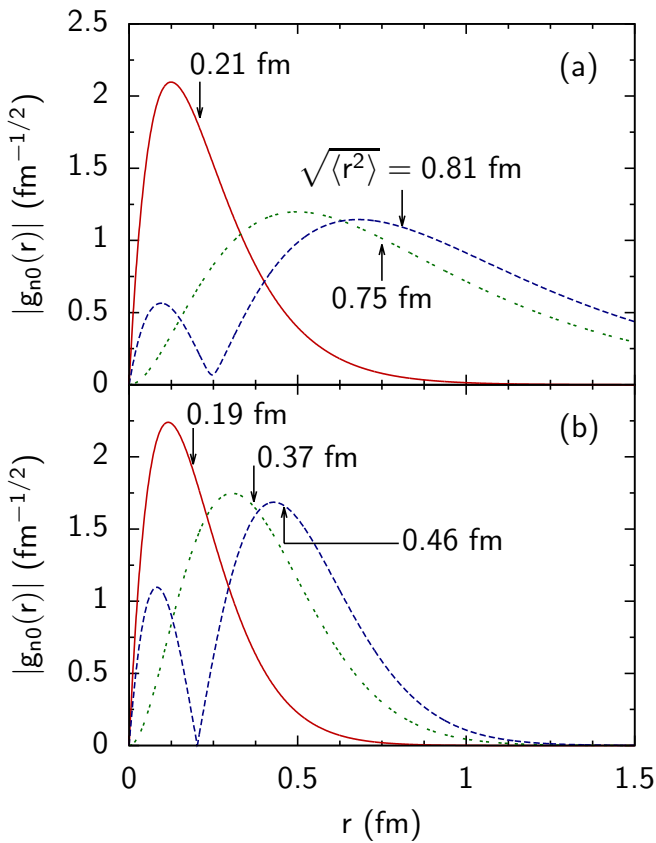}
\caption{
\label{fig1}
(Color online) Radial wave functions of the \us, \cp\ and \uss\ states (solid, dotted, dashed curves, respectively)
calculated in the complex, screened potential (\ref{complex-potential}) for temperatures $T = 200$ MeV (a) and $0$
MeV (b) with effective coupling constant $\alpha_\text{eff} = (4/3) \alpha_s^s = 0.63$, and string tension $\sigma =
0.192$ GeV$^2$. While the rms radius $\sqrt{\langle r^2 \rangle}$ of the \us\ ground state is almost insensitive to temperature changes, it
varies substantially with temperature for the \cp\ and \uss\ states.
}
\end{figure}

In this work we suggest a three step model that considers the $\Upsilon(1S,2S,3S)$ and
$\chi_b(1P,2P)$ states to obtain the suppression of the $\Upsilon(1S,2S,3S)$ states at LHC energies, which is then compared to the experimental results. We successively calculate
\begin{enumerate}
 \item the \bb\ wave functions, and decay widths
for the three processes Debye screening,
collisional damping and gluodissociation \cite{brezinski-wolschin-2012}
 \item  the suppression of the five states considered here within the expanding and cooling fireball
 \item  the feed-down cascade, and the ensuing fraction of dimuon decays, $\muns \rightarrow \mu^+\mu^-$.
\end{enumerate}

Whereas bottomium dissociation due to gluons from the thermal distribution is not possible below $T_c$  where confinement prevails, it does occur above $T_c$ where the color-octet
state of a free quark and antiquark can propagate in the medium. Its significance increases substantially with the
rising gluon density at LHC energies.

The interactions of quarkonia with cold hadronic or nuclear matter are eventually also mediated by gluons and can lead
to dissociation (although not to free quarks and antiquarks in the QGP).  The formalism that we use to calculate the
gluodissociation is in principle also suitable for cold systems if the thermal gluon distribution is replaced by the
appropriate gluon pdfs, but we neglect cold nuclear matter (CNM) effects in the present investigation and focus on
gluodissociation and damping in the hot QGP environment.

Should it turn out that the forthcoming $p$Pb experiments show an unexpected importance of CNM effects in quarkonia suppression, we will have to reconsider them. A related dissociation study with gluon exchanges that is based on open heavy flavor dissociation in the medium was recently performed in \cite{svi12}.

In the midrapidity range $|y|<2.4$ where the CMS measurement \cite{CMS2011a,CMS-2012,CMS2012b} has been performed, the temperature and hence, the thermal
gluon density is high, and causes a rapid dissociation in particular of the \uss\ and \usss\ states, but also of the
\us\ ground state. At larger
rapidities up to the beam value of $y_\text{beam}=7.99$ and correspondingly small scattering angles where the
valence-quark density is high
\cite{mehtar-wolschin-2009}, nonthermal processes would be more important than in the midrapidity region that we are investigating here.

In this work we do not discuss explicitly the production mechanism of the bottomium states, but rather work with initial
populations as deduced from the experimental CMS results in $pp$ at the same center of mass energy \cite{CMS-2012}, and
a distribution according to the number of binary collisions. The final populations of the \uns\ states are measured from
$\mu^+\mu^-$ decays, and we calculate the initial populations through an inverted decay cascade using the CMS 2.76 TeV
data for \uns\ and CDF $p\bar{p}$ data at 1.8 TeV \cite{tevatron2000} for $\chi_b(nP)$.

The calculation of the \bb\ wave functions for five \uns\ and \cnp\ states, and the associated widths \GammaDamp\ of
these states due to collisional damping from a complex potential are considered in the following section. The
calculation of the gluodissociation decay widths \GammaDiss\ for the same states is discussed in Sec. III,
the time evolution of the fireball and subsequent decay cascade in PbPb at
$\sqrt{s_{NN}}$ = 2.76 TeV is considered in Sec. IV. The results are presented in
Sec. V in comparison with the available CMS data at LHC energies, and the conclusions are drawn in Sec. VI.

\section{Bottomium wave functions and collisional damping}

Due to the small relative velocity $v \ll c$ of the bottom quarks in the bound state, \bb\ may be properly described by
the potential NonRelativistic QCD (pNRQCD) approach \cite{caswell-lepage-1986,pineda-soto-1998a,bra00}.
The relevant terms in the pNRQCD action for the \bb-pair read \cite[see
e.g.][]{brambilla-etal-2008,brambilla-etal-2010,brambilla-etal-2011,jacobo-phd}
\begin{align}
 \mathbb{S} &= \int dt d^3R d^3r \, \bigg[ S^\dagger \left( i\del_t + \frac{\Delta_R}{4m_b} + \frac{\Delta_r}{m_b} + \frac{C_F
\alpha_s^s}{r} \right) S \non
  &\hspace{1.5cm}	+ {O^a}^\dagger \left( iD_t + \frac{\Delta_R}{4m_b} + \frac{\Delta_r}{m_b} - \frac{\alpha_s^s}{2N_c r} \right) O^a	\non
  &\hspace{1.5cm}	+ \frac{g}{\sqrt{2N_c}} \vec{r} \vec{E}^a \left( S^\dagger O^a + {O^a}^\dagger S \right) + \dots \bigg],
\label{pNRQCD-action}
\end{align}
with the singlet and octet fields $S$ and $O^a$, the ultra soft color electric field $\vec{E}^a$, the bottom mass $m_b =
4.89$ GeV, the strong coupling constant at the soft scale, $\alpha_s^s = \alpha_s(m_b \alpha_s/2) = 0.48$, and $N_c =
3$, $C_F = (N_c^2-1)/(2N_c) = 4/3$.

This approach leads to a Schr\"odinger equation, with the coulombic, color-singlet potential $V = -C_F \alpha_s^s/r$.
For the treatment of \bb\ in the QGP it is, however, appropriate to make a calculation at finite temperature which
yields for the short-range part of the potential, in the HTL approximation, a complex, screened, coulombic expression
\cite{laine-etal-2007,blaizot-2008} that we use in our phenomenological approach.

The potential does not yet contain the long-range non-perturbative string contribution which causes confinement and vanishes due to screening only at sufficiently high temperature $T>T_c$. Since a consistent derivation is not possible, we parametrize the
long-range part as in \cite{kms88} so that the full singlet potential reads
\begin{align}
 V(r,m_D) &= \frac{\sigma}{m_D} \left( 1 - e^{-m_D r} \right) - \alpha_\text{eff} \left( m_D + \frac{e^{-m_D r}}{r} \right)	\non
&\hspace{0cm}	- i \alpha_\text{eff} T \int\limits_0^\infty \frac{dz \, 2z}{(1+z^2)^2} \left( 1 - \frac{\sin(m_D r z)}{m_D r z}
\right),	\label{complex-potential}	\\
 m_D &= T \sqrt{4\pi \alpha_s^T \left(\frac{N_c}{3} + \frac{N_f}{6}\right)},
\end{align}
with $\alpha_\text{eff} = 4\alpha_s^s/3$, the Debye mass $m_D$, number of flavors in the QGP $N_f = 3$, and the strong
coupling constant evaluated at the HTL energy $2\pi T$, $\alpha_s^T = \alpha_s(2\pi T) \leq 0.50$, respectively. The
absolute values $|g_{nl}(r)|$ of the resulting \bb\ wave functions are shown in Fig.~\ref{fig1}.

The Schr\"odinger equation is now solved for every \bb\ state with the potential (\ref{complex-potential}) for $T \geq
T_c$ up to the dissociation temperature $T_\text{diss}$ above which screening prevents bottomium formation and the
Schr\"odinger equation has no bound states solutions. The dissociation temperatures with the above parameters are $T_\text{diss} \simeq
668$, 217 and 206 MeV for the \us, \uss\ and \cp, respectively: The higher excited states are already dissolved for $T
\gtrsim T_c$.

The imaginary part of the potential causes a decay width \GammaDamp, monotonically increasing with temperature, which
accounts for collisional damping by the plasma particles. \GammaDamp\ is displayed in Fig.~\ref{fig3} together with the
decay width \GammaDiss\ for gluodissociation, which is considered in the next section.

\section{Gluodissociation in the medium}

Due to the high gluon density reached at LHC energies in the mid-rapidity region, gluodissociation is a major process
besides screening and collisional damping that leads to a suppression of $\Upsilon$'s at LHC. Hence we calculate the
gluodissociation cross sections for the \us-\usss, and \cp, \cpp\ states for different lifetimes $t_\text{QGP}$ of the
QGP.

The leading-order dissociation cross section of the \bb\ states through E1 absorption of a single gluon had been derived
by Bhanot and Peskin (BP) \cite{bp79}. From the pNRQCD approach the gluodissociation cross section may be derived from
the dipole interaction term in eq.~(\ref{pNRQCD-action}) describing a singlet-octet transition of the \bb-pair via
emission/absorption of an ultra soft gluon. From this starting point we can easily generalize the approach to include
the effect of our modified potential (\ref{complex-potential}) \cite{dissertation}, and obtain for a bottomium state
$(nl)$

\begin{figure}[tph]
\begin{center}
\includegraphics[width=8.6cm]{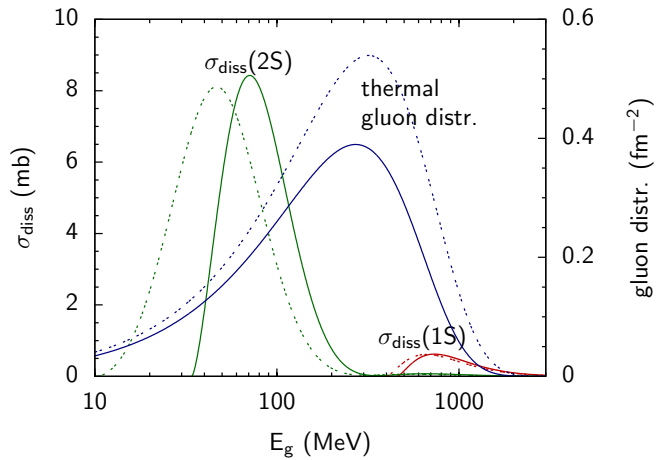}
\caption{\label{fig2}
(Color online) Gluodissociation cross sections $\sigma_\text{diss}(nS)$ in mb (left scale) of the \us\ and \uss\ states
calculated using the screened complex potential for temperatures $T=170$ (solid curves) and $200$ MeV (dotted curves) as
functions of the gluon energy $E_g$. The thermal gluon distribution (right scale; solid for $T = 170$ MeV, dotted for
$200$ MeV) is used to obtain the thermally averaged cross
sections through integrations over the gluon momenta.}
\end{center}
\end{figure}

\begin{align}
 \sigma_{\text{diss},nl}(E_g) &= \frac{2 \pi^2 \alpha_s^u E_g}{(2l+1) N_c^2} \sum\limits_{m=-l}^{l}
\sum\limits_{l'=0}^\infty \sum\limits_{m'=-l'}^{l'}	\non
  &\hspace{0cm}	\cdot \int\limits_0^\infty dq \, | \bra{nlm} \, \hat{\vec{r}} \, \ket{ql'm'} |^2 \delta\left(
E_g+E_{nl}-\frac{q^2}{m_b} \right)	\non
  &\hspace{-1cm}	= \frac{\pi^2 \alpha_s^u E_g}{N_c^2} \sqrt{\frac{m_b}{E_g+E_{nl}}} \, \frac{l |J^{q,l-1}_{nl}|^2 +
(l+1) |J^{q,l+1}_{nl}|^2}{2l+1}, \non
 J^{ql'}_{nl} &= \int\limits_0^\infty dr \, r \, g_{nl}^*(r) h_{ql'}(r), \label{sigma-diss}
\end{align}
with the singlet and octet states $\ket{nlm}$, $\ket{ql'm'}$ and $\alpha_s^u = \alpha_s(m_b \alpha_s^2/2) \simeq 0.59$.
The radial wave function $h_{ql'}$ of the states $\ket{ql'm'}$ is derived from the octet Hamiltonian with the potential
$V_8 = +\alpha_\text{eff}/(8r)$, and the value of q is as determined from energy conservation, $q =
\sqrt{m_b(E_g+E_{nl})}$. The use of the $\delta-$function is an approximation, the actual energy-conserving function in a complex potential acquires a width (Breit-Wigner distribution).

We had originally derived the gluodissociation cross section in \cite{brezinski-wolschin-2012}
independently from the pNRQCD formulation in an approach that was based on a straightforward extension of the
Bhanot-Peskin formulation \cite{bp79} to approximately account for the confining string contribution
\cite{dissertation}.

For vanishing string tension and the corresponding values of the binding energy $E_{nl}$, a pure Coulomb $1S$
wave function, and a simplification in the octet wave function, our expression reduces to the result in \cite{bp79}. Our full result for the \us\ gluodissociation cross section agrees with the result obtained independently by Brambilla et al. in their effective field theory approach \cite{brambilla-etal-2011,jacobo-phd} in the limit discussed in
\cite{brezinski-wolschin-2012}.

\begin{figure}[tph]
\begin{center}
\includegraphics[width=8.6cm]{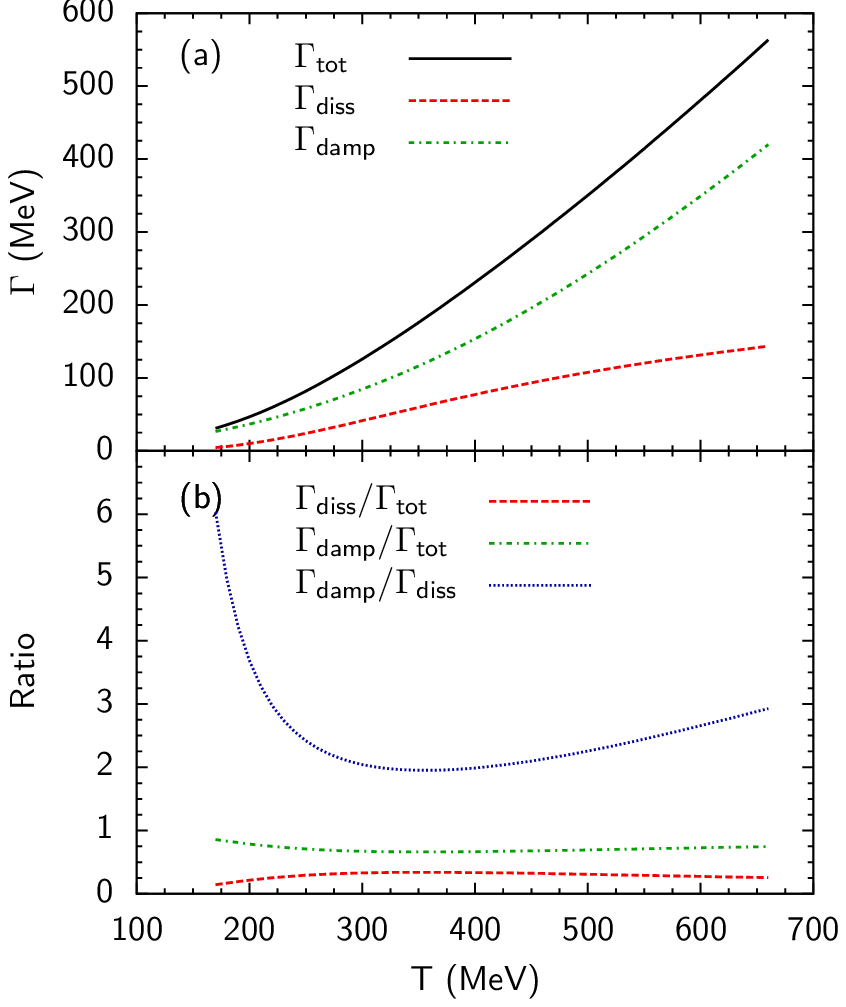}
\caption{\label{fig3}
(Color online) (a) Total decay width \GammaTot\ (solid) and the partial widths for collisional damping
\GammaDamp\ (dot-dashed) and gluodissociation \GammaDiss\ (dashed) for the \us\ state. (b) Ratios of the
partial widths $\mGammaDiss/\mGammaTot$ (dashed) and $\mGammaDamp/\mGammaTot$ (dot-dashed) to the total width, and ratio of the partial widths $\mGammaDamp/\mGammaDiss$ (dotted). While collisional damping is the dominant process in the QGP, gluodissociation can not be neglected for the \us\ state.
}
\end{center}
\end{figure}

To obtain the mean gluodissociation cross section, we average our calculated gluodissociation cross sections over the
Bose-Einstein distribution
function of gluons at temperature $T$, thus assuming that the medium is thermalized, although the heavy \bb\ is not (see
Fig.~\ref{fig2} for the gluon distribution):
\begin{equation}
 \Gamma_{\text{diss},nl} = \frac{g_d}{2\pi^2} \int\limits_0^\infty \frac{dp_g \, p_g^2 \,
\sigma_{\text{diss},nl}(E_g)}{e^{E_g/T}-1},	\label{gamma-diss}
\end{equation}
where $g_d = 16$ is the number of gluonic degrees of freedom. This expression is valid for an idealized case of Upsilons
at rest in a thermal bath of gluons with temperature T. However, produced quarkonia are never strictly at rest, but have
an rms momentum of several GeV.

Studies of Upsilon production show that the mean transverse momentum is about 5-6 GeV/$c$, with corresponding average
velocities of $\langle v \rangle \simeq 0.46-0.54c$, and Lorentz factors $\gamma=1.13-1.18$. This would cause a blue shift in the gluon
distribution that the typical Upsilon sees, corresponding to a distribution with an effective temperature $
T\cdot \sqrt{(1+\langle v \rangle)/(1-\langle v \rangle)}$, and an enhanced dissociation cross section.

On the other hand, the QGP medium also expands with a similar velocity. ALICE has deduced  transverse expansion velocities in 2.76 TeV PbPb in the range 0.5$-0.65 c$ \cite{pr12}; longitudinal velocities are expected to be somewhat larger. Hence the velocity difference that is relevant for the extent of the relativistic Doppler effect is probably small, and we will neglect it in the course of this work, although it would certainly warrant detailed studies.

Taking \GammaDamp\ from the previous section together with
the resulting width from gluodissociation yields the total decay width in the QGP,

\begin{align}
 \Gamma_\text{tot} &= \Gamma_\text{damp} + \Gamma_\text{diss}.	\label{total-decay-width}
\end{align}
\GammaTot\ as well as the partial widths \GammaDamp\ and \GammaDiss\ are displayed in Fig.~\ref{fig3}. The ground-state width from collisional damping is seen to be about twice as large as gluon-induced dissociation in the temperature range 200--400 MeV, such that both processes need to be considered when calculating the total width in the quark-gluon plasma. Whereas damping increases monotonically with temperature, gluodissociation reaches a maximum, and decreases again at very high temperatures beyond 600 MeV due to the diminishing overlap of the thermal gluon distribution and the gluodissociation cross section at large values of T.


\section{Time evolution of the fireball and decay cascade}

The density distribution of the lead ions is modeled by a Woods-Saxon potential with radius $R = 6.62$ fm and
diffuseness $a = 0.546$ fm \cite{nucleusparameters1987}. The number $N_{\mbb}$ of produced \bb-pairs at the point
$(x,y)$ in the transverse plain and impact parameter $b$ is then proportional to the number of binary collisions
$N_\text{col}$ and nuclear overlap $T_{AA}$, $N_{b\bar{b}}(b,x,y) \propto N_\text{coll}(b,x,y) \propto T_{AA}(b,x,y)$.
The initial temperature is parametrized depending on the number of collisions, and Bjorken scaling is used for the time
evolution \cite{QGP},

\begin{figure}[tph]
\begin{center}%
\includegraphics[width=8.6cm]{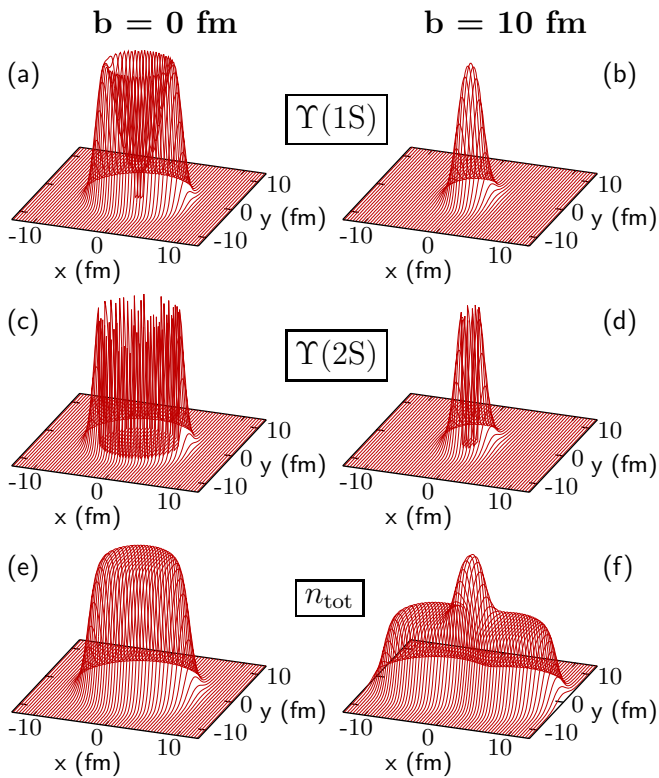}
\caption{\label{fig4} (Color online) Scaled populations (explanation in the text) of \us\ (a,b) and \uss\ (c,d)
which remain after the fireball has cooled, projected on the transverse plane, for $b=0$ fm (left) and $b=10$ fm (right)
at $t_\text{QGP} = 6$ fm/$c$ and $t_F = 0.1$ fm/$c$. The corresponding maximum density of both Pb-nuclei during
the collision is also displayed (e,f). Clearly the \uss\ is suppressed much more efficiently by the dissociation
processes in the QGP. The suppression is stronger in the central regions of the collision where the temperature is
higher.
}
\end{center}
\end{figure}

\begin{figure}[tph]
\begin{center}%
\includegraphics[width=8.6cm]{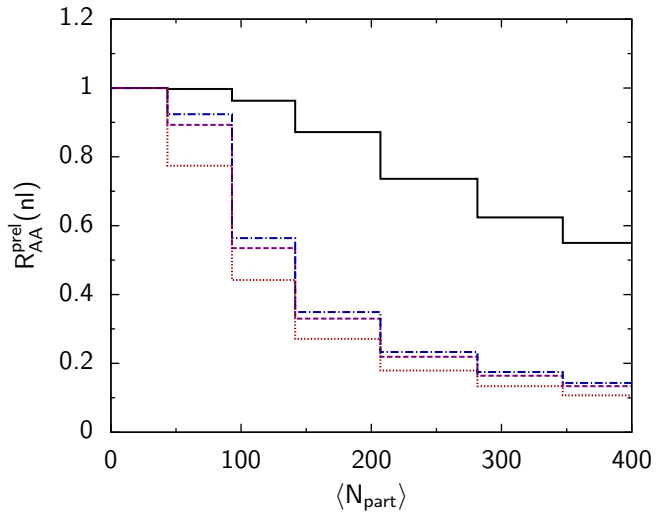}
\caption{\label{fig5}
(Color online) Preliminary suppression factors \Rprel(nl) from eq.~(\ref{preliminaryRAA}) as functions of centrality
for the different bottomium states \us\ (solid), \uss\ (dash-dotted) \cp\ (dashed),  and higher excited states (dotted)
for the formation time $t_{F} = 0.1$ fm/$c$ and QGP lifetime $t_\text{QGP} = 6$ fm/$c$.
}
\end{center}
\end{figure}

\begin{align}
 T(b,t,x,y) &= T_c \, \frac{T_{AA}(b,x,y)}{T_{AA}(0,0,0)} \left( \frac{t_\text{QGP}}{t} \right)^{1/3},
\end{align}
where $t_\text{QGP}$ is the maximum lifetime of the quark-gluon plasma.

In view of the principle lack of knowledge about a more complete understanding of the initial
stages of the collision and the associated entropy production, as well as the unknown relation between the number of
binary collisions (or, in a different formulation, the number of participants, or a mixture between the two) and the initial gluon content, the above assumption for the space and time dependence of the temperature is certainly open for improvement. Changing the space-time dependent ansatz for the temperature will, in particular, lead to a different centrality dependence and magnitude of the calculated quarkonium suppression.

We define a preliminary suppression factor
\Rprel, which accounts only for the \bb\ suppression due to the three processes Debye screening, collisional damping and
gluodissociation,
\begin{align}
 \mRprel &= \frac{\int d^2b \int dxdy \, T_{AA}(b,x,y) \, e^{-\int_{t_F}^\infty dt \, \Gamma_\text{tot}(b,t,x,y)}}{\int
d^2b \int
dxdy \, T_{AA}(b,x,y)}.	\label{preliminaryRAA}
\end{align}
The numerator of eq.~(\ref{preliminaryRAA}) is proportional to the number of \bb\ bound states which have survived from
their formation time $t_{F}$ until the fireball has cooled below the critical temperature $T_c$, where the decay width
\GammaTot\ is set to vanish. The integrand in the numerator of \Rprel,
\begin{align}
 T_{AA}(b,x,y) \, e^{-\int\limits_{t_F}^\infty dt \, \Gamma_\text{tot}(b,t,x,y)},
\end{align}
is displayed in Fig.~\ref{fig4} for the \us\ and \uss\ states for a central ($b = 0$ fm) and a peripheral collision ($b
= 10$ fm). The total density of the PbPb-system in the moment of the collision where the nuclei pass through each other
is also displayed. Clearly the \uss\ is suppressed much more efficiently than the more stable \us. Also one should note
the action of Debye screening which forbids the formation of bound \bb\ states at sufficiently high temperatures and
thus changes the shape of the surface from bell-shape (peripheral) into volcano-like (central).

Results for the preliminary suppression factor of all five states for formation time $t_{F} = 0.1$ fm/$c$ and
$t_\text{QGP} = $ 6 fm$/c$ are presented in Fig.~\ref{fig5}. For the excited states \cpp\ and \usss\ and higher
excitations there exist no bound states for $T \geq T_c$ so that \Rprel\ is equal for all these states.

\begin{figure}[tph]
\begin{center}
\includegraphics[width=8.6cm]{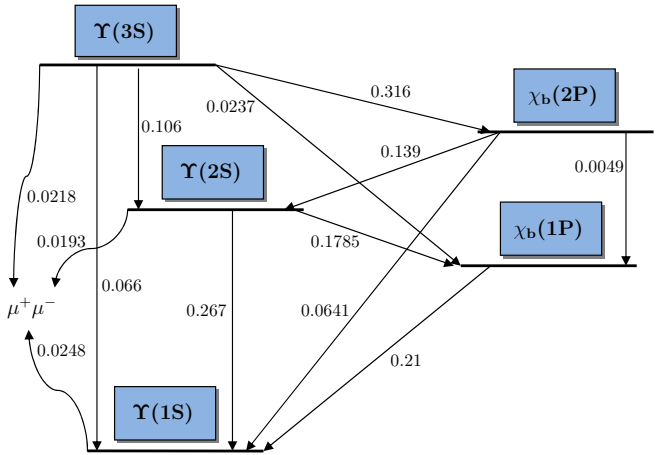}
\caption{\label{fig6}(Color online) Branching ratios for decays within the bottomium family \uns\ and \cnp\ and into
$\mu^\pm$-pairs according to \cite{pdg2012}.}
\end{center}
\end{figure}

Now that we have calculated the suppression during the evolution of the fireball we have to consider the feed-down of
the remaining \bb\ population to calculate the fraction of decays into dimuon pairs, $\muns \rightarrow \mu^+ \mu^-$ .
Fig.~\ref{fig6} displays the decays within the \bb\ family and into dimuon pairs that are measured. Considering first
the processes inside the fireball and then performing the decay cascade as a subsequent step is justified by the very
different time scales involved. At the LHC the fireball has cooled within less than $\lesssim$ 10 fm/$c$, while the
subsequent decays take place on time scales $\sim 10^{3}$ fm/$c$.

Let us denote \bb\ states by $I=(nl)$ and $(\mathcal{C}_{IJ})$ ($I \leq J$) the branching ratio of state $J$ to decay
into state $I$ including all indirect decays with intermediate \bb\ states. The initial and final \bb\ numbers of state
$I$, $N^i(I)$ and $N^f(I)$ in pp and PbPb collisions are then connected by
\begin{align}
 N^f_{pp}(I) &= \sum_{I \leq J} \mathcal{C}_{IJ} N^i(J),	\non
 N^f_\text{PbPb}(I) &= \sum_{I \leq J} \mathcal{C}_{IJ} N^i(J) \mRprel(J).
\end{align}
Further we define the number of \uns\ states that decay into dimuon pairs
\begin{align}
 N^f_{\mu^\pm}(nS) = \mathcal{B}(nS\rightarrow \mu^\pm) N^f_\text{PbPb}(nS),
\end{align}
where $\mathcal{B}(nS\rightarrow \mu^\pm)$ is the corresponding branching ratio.

We take $N^f_{\mu^\pm}(nS)$ from the
2012 CMS data \cite{CMS-2012} and consider that 27.1\% and 10.5\% of the \us\ population comes from \cp\ and \cpp\ decays,
respectively \cite{tevatron2000}. Since these CDF results are obtained from $p\bar{p}$ collisions at 1.8 TeV
with a transverse momentum cut $p_T^{\Upsilon}>8.0$ GeV/$c$,
it would be desirable to confirm the \us\ populations from $\chi_b$ decays in new $pp$ measurements at 2.76 TeV, which
are not yet available.

 The initial populations are then obtained in units of $\mathcal{B}(nS\rightarrow
\mu^\pm) N^f_{pp}(1S)$ as $N^i(1S) = 16.2$, $N^i(1P) = 43.7$, $N^i(2S) = 20.3$, $N^i(2P) = 45.6$, $N^i(3S) = 18.8$. The
final suppression factor is now simply calculated as $R_{AA}(nS) = N^f_{\mu^\pm}(nS)/N^f_{pp}(nS)$ or
\begin{align}
 R_{AA}(nS) &= \mathcal{B}(nS\rightarrow \mu^\pm) \frac{\sum_{nS \leq J} \mathcal{C}_{IJ} N^i(J) \mRprel(J)}{\sum_{nS
\leq J} \mathcal{C}_{IJ} N^i(J)}.	\label{finalRAA}
\end{align}
\begin{figure}[t]
	\centering
\includegraphics[width=8.6cm]{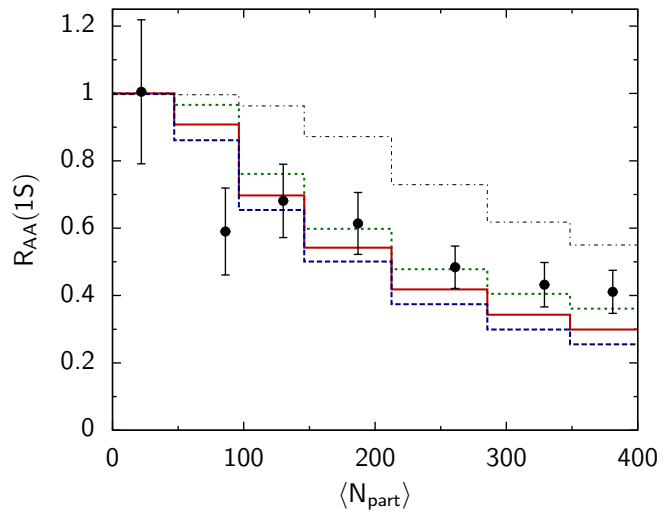}
\caption{\label{fig7}(Color online) Suppression factor $R_{AA}$ for the \us\ ground state calculated for $2.76$ TeV PbPb-collisions
from screening, collisional damping, gluodissociation and reduced feed-down using three QGP lifetimes $t_\text{QGP}=4$, 6, 8
fm/$c$ (dotted, solid and dashed line respectively) for the
centrality bins 50--100\%, 40--50\%, 30--40\%, 20--30\%, 10--20\%, 5--10\%, 0-5\%. The dash-dotted upper line is the preliminary suppression factor \Rprel(1S) ($t_\text{QGP} = 6$ fm/$c$) without reduced feed-down. The corresponding CMS data
\cite{CMS-2012} are in good agreement with the model results for the \us\ state.
}
\end{figure}

\begin{figure*}[t]
	\centering
\includegraphics[width=12.8cm]{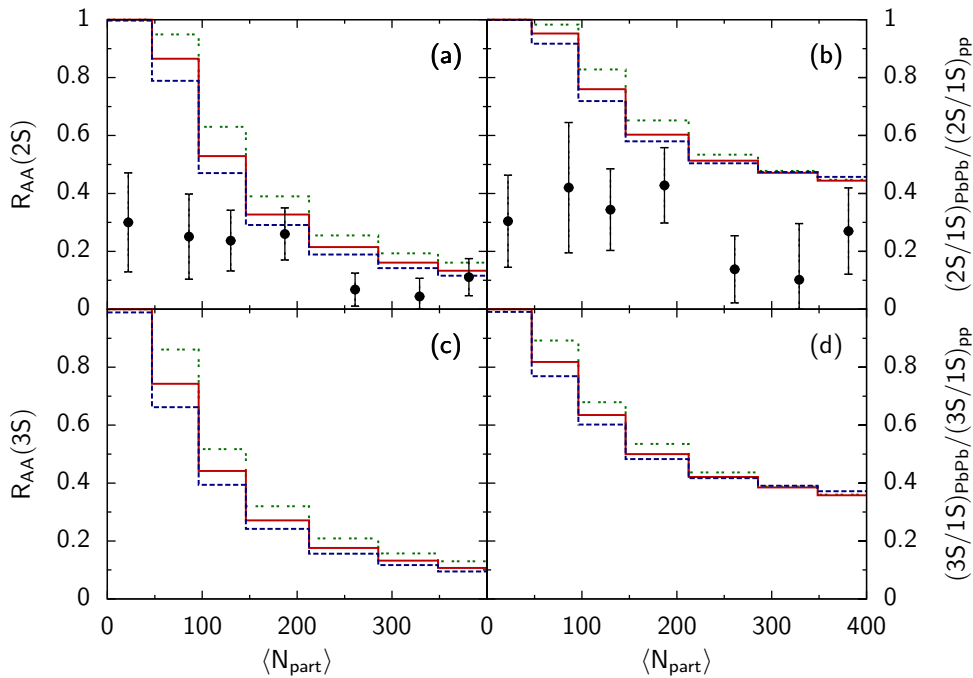}
\caption{\label{fig8}(Color online) Suppression factors $R_{AA}$ for the \uss, \usss\ states (a,c)
and the double ratios ($n$S/1S)$_\text{PbPb}$/($n$S/1S)$_{pp}$ for $n = 2,3$ (b,d) calculated for
$2.76$ TeV PbPb-collisions from screening, collisional damping, gluodissociation and feed-down using three QGP lifetimes
$t_\text{QGP}=4$, 6, 8 fm/$c$ (dotted, solid and dashed line respectively) for the centrality bins 50--100\%, 40--50\%,
30--40\%, 20--30\%, 10--20\%, 5--10\%, 0--5\% (left to right). The corresponding CMS results \cite{CMS-2012} for
the \uss\ state show significantly more suppression, in particular, in the peripheral region.
}
\end{figure*}

\begin{table}[bph]
\caption{\label{tab:min-bias-results}
Calculated minimum bias results for different $t_\text{QGP}$ and $t_F = 0.1$ fm/$c$ compared to the CMS results
\cite{CMS-2012} with statistical and systematic error bars, respectively. The $R_{AA}(1S)$ is in good agreement
with experiment, but the results for the excited states allow for additional suppression mechanisms.
}
\vspace{1em}
\begin{tabular}{ccccc}	\hline\hline
 $t_\text{QGP}$ (fm$/c$) & 4 & 6 & 8 & CMS data \cite{CMS-2012}	\\	\hline
 $R_{AA}(1S)$ & 0.51    & 0.46 & 0.41 & $0.56 \pm 0.08 \pm 0.07$	\\
 $R_{AA}(2S)$ & 0.33   & 0.28 & 0.25 & $0.12 \pm 0.04 \pm 0.02$	\\
 $R_{AA}(3S)$ & 0.28  & 0.24 & 0.22 & $0.03 \pm 0.04 \pm 0.01$	\\
 $\frac{(2S/1S)_\text{PbPb}}{(2S/1S)_{pp}}$ & 0.63 & 0.61 & 0.61 & $0.21 \pm 0.07 \pm 0.02$
\\
 $\frac{(3S/1S)_\text{PbPb}}{(3S/1S)_{pp}}$ & 0.54 & 0.52 & 0.52 & $0.06 \pm 0.06 \pm 0.06$
\vspace{1mm}	\\
\hline\hline
\end{tabular}
\end{table}

\section{Results}

We present the results for screening and collisional damping derived from the solutions of the Schr\"odinger equation
with the potential eq.~(\ref{complex-potential}), and the widths for gluodissociation as derived from eq.
(\ref{gamma-diss}). The total decay widths \GammaTot\  are then inserted into a dynamic
calculation for the fireball evolution to calculate preliminary suppression factors, eq.~(\ref{preliminaryRAA}).

Subsequently, the bottomium states pass through a decay cascade (see Fig.~\ref{fig6}) so that the higher excited states
feed the lower lying states to yield the final suppression factor eq.~(\ref{finalRAA}).

Our results for the suppression of the \us\ state in PbPb relative to $pp$ are shown in Fig.~\ref{fig7} for three
different QGP lifetimes $t_\text{QGP} = 4$, 6, 8 fm/$c$ as functions of centrality (number of participants). When
comparing with our result from the preliminary suppression factor (upper dotted step function), it is evident that the
consideration of the feed-down cascade is essential for modeling the suppression.

The CMS data point \cite{CMS-2012} at 40-50\% centrality violates the monotonic increase of the suppression with
centrality, but is consistent with the other points within statistical and systematic error bars. Hence, the
calculated suppression is in very good agreement with the CMS data for the \us\ ground state. This is also true for minimum
bias (centrality integrated) results, which are shown in Table~\ref{tab:min-bias-results}.

There is a considerable dependence of the calculated suppression factors on the Upsilon formation time.
Generally shorter formation times lead to more suppression in the QGP, because the dissociation processes start to act
at a higher initial temperature and hence, are more efficient. Typical results for the suppression of \us\ in minimum
bias collisions with gluodissociation and damping for a QGP lifetime of 6 fm/c  are $R_{AA}(1S) = 0.74, 0.63$ and 0.45
for $t_F = 1$, 0.5 and 0.1 fm/$c$, respectively.

Our results for the suppression of the \uss\ and \usss\ states in PbPb relative to $pp$ are shown in
Fig.~\ref{fig8}
(left column) for three different QGP lifetimes $t_\text{QGP} = 4$, 6, 8 fm/$c$ as functions of centrality. The double
ratios with respect to the \us\ state in PbPb and $pp$ are displayed in the right column of Fig.~\ref{fig8}, with CMS
data \cite{CMS-2012} included for the \uss\ state. The suppression found experimentally for the \uss\ state is much more
pronounced than in the calculation, in particular, for the three more peripheral data points.

It appears to be very difficult for theoretical models to obtain such a huge suppression of the \uss\ state in
peripheral collisions. Indeed, other approaches such as \cite{shk12,ezr12,str12} also find that the  \uss\ suppression
factor rises towards 1 for peripheral collisions. As a consequence of the disagreement with the centrality-dependent data,
our minimum-bias results of  Table~\ref{tab:min-bias-results} also disagree substantially for the \uss\ and \usss\
states.

The reason for the disagreement will probably be cleared up once more precise $pp$ reference data at 2.76 TeV become available in the future. It is, however, also conceivable that additional suppression mechanisms not considered in this work play a role for the \uss\ and \usss\ states.

\section{Conclusion}

We have formulated a three-step model for the suppression of the bottomium states \uns\ in the quark-gluon plasma that
is formed in PbPb collisions at LHC energies. Due to its stability against screening up to very high temperatures, the
\us\ state is a particularly suitable probe for the relevance of gluodissociation, collisional damping, and reduced
feed-down.

We find that gluodissociation of the \us\ state is sizeable \cite{brezinski-wolschin-2012} due to the strong overlap of
the \us\ gluodissociation cross section with the thermal gluon distribution.  In the temperature region 200--400 MeV, both gluodissociation and collisional damping are found to be important.

The observed suppression factor $R_{AA}(1S) = 0.56$ in minimum-bias PbPb collisions \cite{CMS-2012} is essentially due
to gluodissociation and damping of the \us\ state, and to the melting and dissociation of the excited states: The
excited states -- in particular, the \cnp\ states -- partially feed the \us\ state in $pp$, $p\bar{p}$ and $e^+e^-$
collisions, and their melting and dissociation in the quark-gluon plasma substantially reduces the feed-down in PbPb
collisions at LHC energies.

The calculated \us\ suppression factor as function of the collision centrality is
indeed in very good agreement with the CMS data if the modification of the feed-down cascade in PbPb as compared to $pp$ is taken into account.

Different from the \us\ ground state, the excited states -- and in particular, the \uss\ and \usss\ states that are
observed in the CMS experiment -- are already suppressed through screening to a much larger extent than the ground
state, so that the contributions from damping and gluodissociation are less important here. The dissolution of the
excited states in the quark-gluon plasma causes the substantial feed-down reduction that is one of the three main
reasons for the ground-state suppression.

From our calculations it appears that there may be additional causes for the suppression of the excited states, such as cold nuclear matter (CNM) effects -- although these should essentially cancel out in the double ratios that are shown in Fig.~\ref{fig8}. It is conceivable that CNM-effects will be constrained in forthcoming $p$Pb measurements at the LHC. Compared to the present CMS experimental results for the suppression of the \uss\ and \usss\ states in PbPb \cite{CMS-2012}, our calculated $R_{AA}$ values are substantially too large, in particular, in peripheral collisions.

Our phenomenological approach to Upsilon suppression in PbPb collisions at LHC energies thus yields a straightforward description of the ground state suppression due to gluodissociation, damping, and reduced feed-down, although there are caveats related to various model assumptions. Screening is unimportant for the \us\ state. For the excited states the model reveals substantial screening effects and -- together with the other dissociation processes that we consider -- larger suppression than for \us\, but it disagrees quantitatively with the current CMS data regarding the centrality dependence.  Hence there is considerable room for future improvement.

\begin{acknowledgments}
This work has been supported by the IMPRS-PTFS and the ExtreMe Matter Institute EMMI.
\end{acknowledgments}

\bibliographystyle{prsty}
\bibliography{references}

\end{document}